\documentclass[12pt,reprint,sort&compress, superscriptaddress, nofootinbib]{revtex4-1} 
\usepackage{graphicx}  
\usepackage{dcolumn}   
\usepackage{bm}        
\usepackage{amssymb}   
\usepackage{hyperref}
\usepackage[UKenglish]{babel} 
\usepackage{color}
\usepackage{lipsum}

\newcommand*{\no}{\noindent}
\newcommand*{\bea}{\begin{eqnarray}}
\newcommand*{\eea}{\end{eqnarray}}
\newcommand*{\be}{\begin{equation}}
\newcommand*{\ee}{\end{equation}}

\newcommand*{\pref}[1]{(\ref{#1})}

\newcommand*{\nn}{\nonumber}

\newcommand*{\tr}{\mathrm{tr}}

\newcommand{\bma}{\begin{pmatrix}}
\newcommand{\ema}{\end{pmatrix}}

\begin{document}

\title{Predicting the singlet vector channel in a partially broken gauge-Higgs theory}
%
%
\author{A.~Maas}
\email{axel.maas@uni-graz.at}
\affiliation{Institute of Physics, NAWI Graz, University of Graz, Universit\"atsplatz 5, 8010 Graz, Austria}

\author{P.~T\"orek}
\email{pascal.toerek@uni-graz.at}
\affiliation{Institute of Physics, NAWI Graz, University of Graz, Universit\"atsplatz 5, 8010 Graz, Austria}

\vskip 0.25cm
\date{\today}


\begin{abstract}
We study a toy version of a grand-unified theory on the lattice: An $SU(3)$ gauge theory, which experiences a Brout-Englert-Higgs effect due to a single Higgs field in the fundamental representation. This yields a perturbative breaking pattern $SU(3) \rightarrow SU(2)$. We investigate the singlet vector channel, finding a non-degenerate and massive ground state. This is in contradistinction to the perturbative prediction of three massless and five massive vector states, even though the correlation functions of the gauge bosons exhibit a weak-coupling behavior, being almost tree-level-like. However, a combination of perturbation theory with the Fr\"ohlich-Morchio-Strocchi mechanism, and thus passing to gauge-invariant perturbation theory, allows to predict the physical spectrum in this channel.
\end{abstract}

\maketitle

\section{Introduction}
Particle physics is successfully described by gauge theories. Within these theories experimentally observable states must be gauge-invariant. This is not an issue in confining theories, like QCD, but is potentially so in the electroweak sector \cite{'tHooft:1979bj,Banks:1979fi,Frohlich:1980gj}: The $W$/$Z$-bosons and the Higgs are not gauge-invariant states. Why is perturbation theory then so successful by using these particles as if they were observable states?

This apparent contradiction is resolved in the standard model by the Fr\"ohlich-Morchio-Strocchi (FMS) mechanism \cite{Frohlich:1980gj,Frohlich:1981yi}. It shows that, in presence of the Brout-Englert-Higgs (BEH) effect, the spectrum of the physical states and the elementary states\footnote{Elementary states are the ones described by the gauge-dependent elementary fields in the Lagrangian, in our case the one in \pref{eq:eccontaction}.} coincide. However, the multiplet structure of the gauge sector is thereby traded for a multiplet structure in the custodial symmetry. But because the gauge group and the custodial group coincide, this yields the same degeneracy pattern. The dynamical condition is that the Higgs fluctuations are small. This is the case in the standard model, and consequently lattice simulations supported this picture \cite{Maas:2012tj,Maas:2013aia}.

However, the applicability of the FMS mechanism relies on the special structure of the standard model and the smallness of the fluctuations. It is therefore not guaranteed to work also in beyond-the-standard-models scenarios \cite{Maas:2015gma}, though it may \cite{Maas:2015gma,Maas:2016qpu}. The aim of this paper is to construct a structural counter-example, i.e. one in which, even if the Higgs fluctuations are small, the degeneration pattern of gauge-invariant states cannot be the one of the gauge-dependent (elementary) states. This situation can be expected in theories with grand-unified-theory-inspired patterns, when the gauge group is larger than the global symmetry group \cite{Maas:2015gma,Torek:2015ssa}. Indeed, it will be seen that only using the FMS mechanism to extend perturbation theory to so-called \cite{Seiler:2015rwa} gauge-invariant perturbation theory this is again possible.

\section{The toy model}

To this end, we consider as a toy model an $SU(3)$ gauge theory with a single Higgs field $\phi$ in the fundamental representation. This theory can exhibit a BEH effect. Its only global symmetry is a U(1) symmetry, acting like a baryon number for the Higgs field\footnote{This symmetry had been overlooked in \cite{Torek:2015ssa}, but plays no role for the argument there.}.

The (Euclidean) Lagrangian of this model is given by
\begin{eqnarray}
\mathcal{L}(x) &= \big(D_\mu\phi(x)\big)^\dagger\big(D_\mu\phi(x)\big)+\mu^2\phi(x)^\dagger\phi(x)\nonumber \\
&-\frac{\mu^2}{2v^2}\left(\phi(x)^\dagger\phi(x)\right)^2+\frac{1}{2}\text{tr}\left[W_{\mu\nu}(x)^2\right]\;,
\label{eq:eccontaction}
\end{eqnarray}
\no where the $W_\mu$ are the gauge fields, $W_{\mu\nu}$ is the usual field-strength tensor, and $D_\mu$ the usual covariant derivative, the later two involving the gauge coupling $g$. The potential has a minimum at $\phi^\dagger\phi=v^2$, and the parameter $\mu^2$ then tunes the Higgs mass. We note here that this theory, as any theory with a BEH effect due to an elementary scalar, is potentially trivial, and may therefore not have a continuum limit in lattice calculations or have the cutoff removed in continuum calculations. We make the usual assumption that this does not affect the physics at low energies beyond corrections suppressed by inverse powers of the cutoff, see \cite{Hasenfratz:1986za} for a detailed discussion.

\section{Analytical analysis of the spectrum}

The usual perturbative construction \cite{Bohm:2001yx,Torek:2015ssa} yields a breaking pattern where the gauge group breaks to $SU(2)$. This yields five massive, the four lighter ones degenerate, and three massless gauge bosons as well as one massive Higgs boson. The relative sizes of the masses depend on the parameters. Whether the unbroken subgroup shows a Coulomb-like structure or confinement plays no role in the following, if the perturbative picture is correct and the infrared dynamics of this sector decouples from the massive states. Indeed, this will be found.

Given that a conflict between the physical spectrum and the perturbative prediction is expected in the vector channel \cite{Torek:2015ssa}, we will concentrate here on the $J^P=1^-$ singlet channel, i.e. not carrying any global U(1) charge. In the diction of QCD, this corresponds to a vector meson. We will investigate further channels elsewhere \cite{Maas:unpublishedtoerek}.

The basic gauge-invariant operator, which we use to investigate this channel, is
\begin{equation}
O_\mu(x) = i(\phi^\dagger D_\mu \phi)(x)\;.
\label{eq:op}
\end{equation}
\no As a composite, gauge-invariant operator, it will not create an asymptotic single-particle state in perturbation theory. To obtain a prediction for its mass spectrum, we apply the FMS prescription \cite{Frohlich:1980gj,Frohlich:1981yi} to the correlator $\langle O(x)O^\dagger(y)\rangle$. This is done by fixing to (minimal) 't Hooft-Landau gauge and expanding the Higgs field around its vacuum expectation value, i.e. $\phi_i(x) = v~\delta_{i,3} + \eta_i(x)$. This yields
\begin{equation}
\langle O_\mu(x) O^\dagger_\mu(y)\rangle = v^4\langle W_\mu^8(x)W_\mu^8(y)\rangle + \mathcal{O}(\eta W/v)\;.
\label{eq:fmsexp}
\end{equation}
\no $W^8$ is the non-degenerate and heaviest massive gauge boson field. Neglecting the higher order terms, this implies that both correlators should have the same mass poles, and thus the same mass. This is the same relation which has been found for the electroweak sector of the standard model \cite{Frohlich:1980gj,Frohlich:1981yi,Maas:2012tj,Maas:2013aia}, with one important difference: The left-hand-side is a singlet, and therefore is expected to have only a single ground-state mass pole. On the right-hand-side also only a single pole contributes. Thus, only a single massive particle is predicted by the FMS mechanism, in contrast to the perturbative prediction. All more involved operators we have considered \cite{Torek:2015ssa,Maas:unpublishedtoerek} only yielded a four-point function as leading order on the right-hand side, and thus only multi-particle states. This analytic result requires confirmation.

\section{Lattice analysis of the spectrum}
As bound states are described by composite operators, which are gauge-invariant, they are not accessible using perturbative means. 
To investigate and confirm the results above, we use here lattice gauge theory as a non-perturbative tool, following closely \cite{Maas:2012tj,Maas:2013aia,Maas:2014pba}.

The lattice action of this model is given by \cite{Montvay:1994cy}
\bea
S[U,\phi]&=&\sum_x\bigg[\phi(x)^\dagger\phi(x) + \lambda\left(\phi(x)^\dagger\phi(x)-1\right)^2 \nn\\
&&-\kappa\sum\limits_{\mu=\pm 1}^{\pm 4}\phi(x)^\dagger U_\mu(x)\phi(x+\hat{\mu})\nn\\
&&+\frac{\beta}{3}\sum_{\mu<\nu}\text{Re tr}[\openone-U_{\mu\nu}(x)]\bigg]\;,
\eea
\no where the link variables $U_\mu$ describe the gauge bosons via $W_\mu(x)=\frac{1}{2ia}\left.(U_\mu(x)-U_\mu(x)^\dagger)\right|_{\text{traceless}}$ and the plaquette $U_{\mu\nu}(x)=U_\mu(x)U_\nu(x+\hat{\mu})U_\mu(x+\hat{\nu})^\dagger U_\nu(x)^\dagger$ the field-strength tensor squared. The three lattice parameters $\beta$, $\kappa$ and $\lambda$ are related to the continuum ones by
\begin{equation}
\beta = \frac{6}{g^2}\;,\, a^2\mu^2 = \frac{1-2\lambda}{\kappa}-8\;, v^2 = -a^2\mu^2\frac{\kappa^2}{2\lambda}\;,
\label{eq:params}
\end{equation}
\no where $a$ is the lattice spacing. The simulations have been performed using a standard multi-hit-Metropolis algorithm\footnote{This algorithm attempts to update a gauge link $5$ times before continuing to the next link.}, where $300+10L$ initial configurations for thermalization and $3L$ configurations in between measurements for decorrelation have been dropped, with $L$ being the lattice size. The integrated autocorrelation time of the plaquette is $\tau_{\text{int}}\approx 1/2$, i.e. close to the minimal value. Thus, no significant correlations between consecutive measurements of gauge-invariant observables are observed. Also, several independent runs have been performed to reduce correlations further.

\begin{figure}[!t]
\centering
\includegraphics[width=0.5\textwidth]{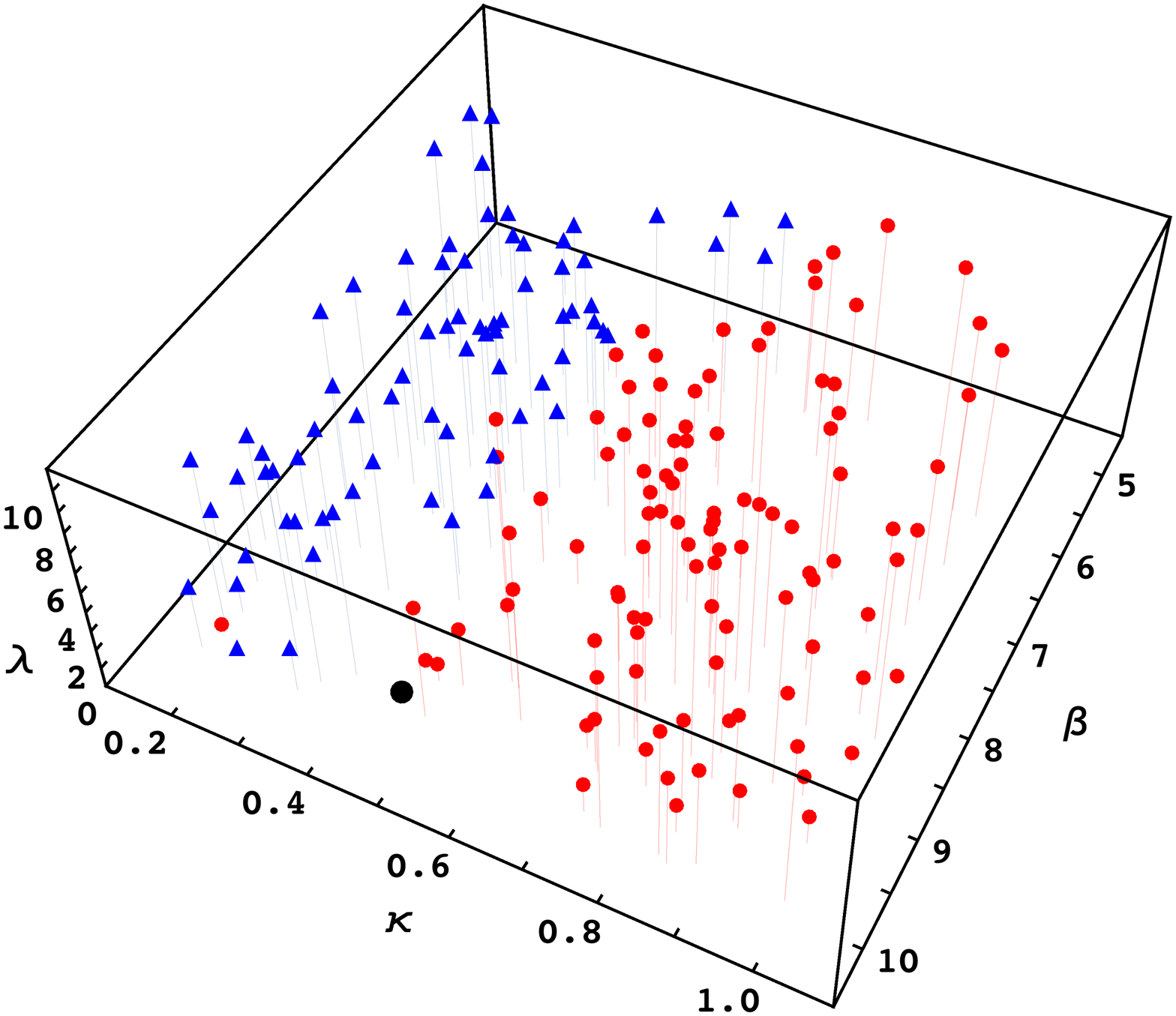}\\
\hspace*{-2.2cm}
\includegraphics[width=0.5\textwidth]{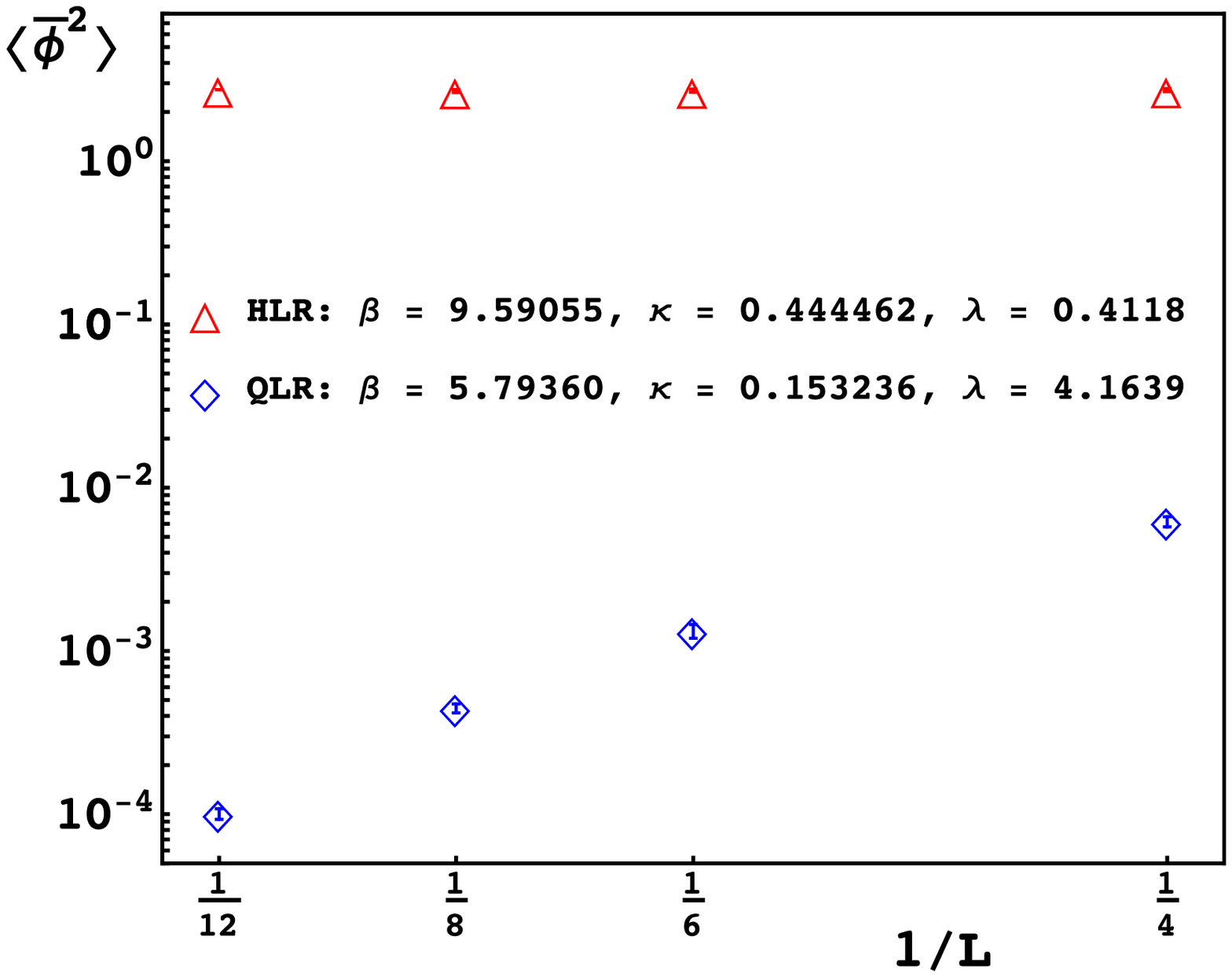}
\caption{The top panel shows the 'phase' diagram of the theory. The blue triangles show no BEH effect in 't Hooft-Landau gauge, while the red points do\footnote{The red point in the lower left corner in the 'blue region' shows indeed a BEH effect. This effect has also been observed in \cite{Maas:2013aia} for large values of $\beta$ and low values of $\lambda$.}. The big black point denotes our primary simulation point at $\beta=9.59055$, $\kappa=0.444462$, $\lambda=0.4118$. The lower panel shows examples for \pref{eq:caudy} as a function of volume, the Higgs case being the simulation point. This demonstrates how the different 'phases' are identified.}
\label{fig:pd}
\end{figure}

Because the gauge group is not fully broken, the arguments of \cite{Osterwalder:1977pc,Fradkin:1978dv} do not apply, and this theory may or may not have separated phases and a possibly rich phase structure. Since we are only interested in a situation with a perturbatively accessible BEH effect \cite{Frohlich:1980gj,Maas:2012ct,Maas:2012tj,Maas:2013aia}, we scanned the phase diagram using the quantity \cite{Caudy:2007sf}
\begin{equation}
\langle\bar{\phi}^2\rangle = \left\langle \left|\frac{1}{V}\sum_x\phi(x)\right|^2 \right\rangle\;,
\label{eq:caudy}
\end{equation}
\no in 't Hooft-Landau gauge, where $V=L^4$ denotes the volume of our hypercubic lattices of size $L$. In the presence of the BEH effect this quantity is finite in the infinite-volume limit and otherwise decays as an inverse power of the volume \cite{Caudy:2007sf}. Examples of this behavior are shown in Figure \ref{fig:pd}.

To scan the phase diagram quickly we performed simulations for $8^4$, $10^4$ and $12^4$ lattices for randomly distributed parameters $\beta$, $\kappa$ and $\lambda$, fixing the gauge as described in \cite{Maas:2013aia} using stochastic overrelaxation\footnote{We used 1000 gauge-fixed configurations for this analysis.}. We furthermore rotate by a global gauge transformation the Higgs expectation value in the real 3 direction afterwards to comply with the 't Hooft-Landau gauge condition. This yielded the 'phase' diagram shown in Figure \ref{fig:pd}, though we neither did check nor need in the following whether the two regions are separated by a genuine phase transition and are really distinct phases. For large values of $\kappa$, i.e. large negative tree-level masses (see Equation \pref{eq:params}) and thus a steep potential, we observe a BEH effect in this gauge, whereas for relatively small values of $\kappa$ there is mainly no BEH effect present. As our simulation point we choose
$\beta=9.59055$, $\kappa=0.444462$, $\lambda=0.4118$, close to the boundary between both regions. This choice is motivated by the $SU(2)$ results where close to the boundary the largest cutoffs, i.e. the smallest lattice spacings, have been found \cite{Maas:2014pba}.

To test whether the system really behaves perturbatively and to test the FMS prediction, i.e. Equation \pref{eq:fmsexp}, the gauge-boson propagator $D^{bc}(p^2)=\langle W_\mu^b(-p)W_\mu^c(p)\rangle$ will be necessary. It is determined in the same way as in \cite{Maas:2013aia}. The propagator is diagonal in this gauge, i.e. $D^{bc}(p^2)=\delta^{bc}D^b(p^2)$, and the averaged propagator in the unbroken $SU(2)$ subgroup ($b=1-3$) and the broken coset ($b=4-7$ are degenerate and $b=8$ is the most massive one \cite{Torek:2015ssa}) are determined separately.

\begin{figure}[!h]
\centering
\hspace*{-1cm}
\includegraphics[width=0.5\textwidth]{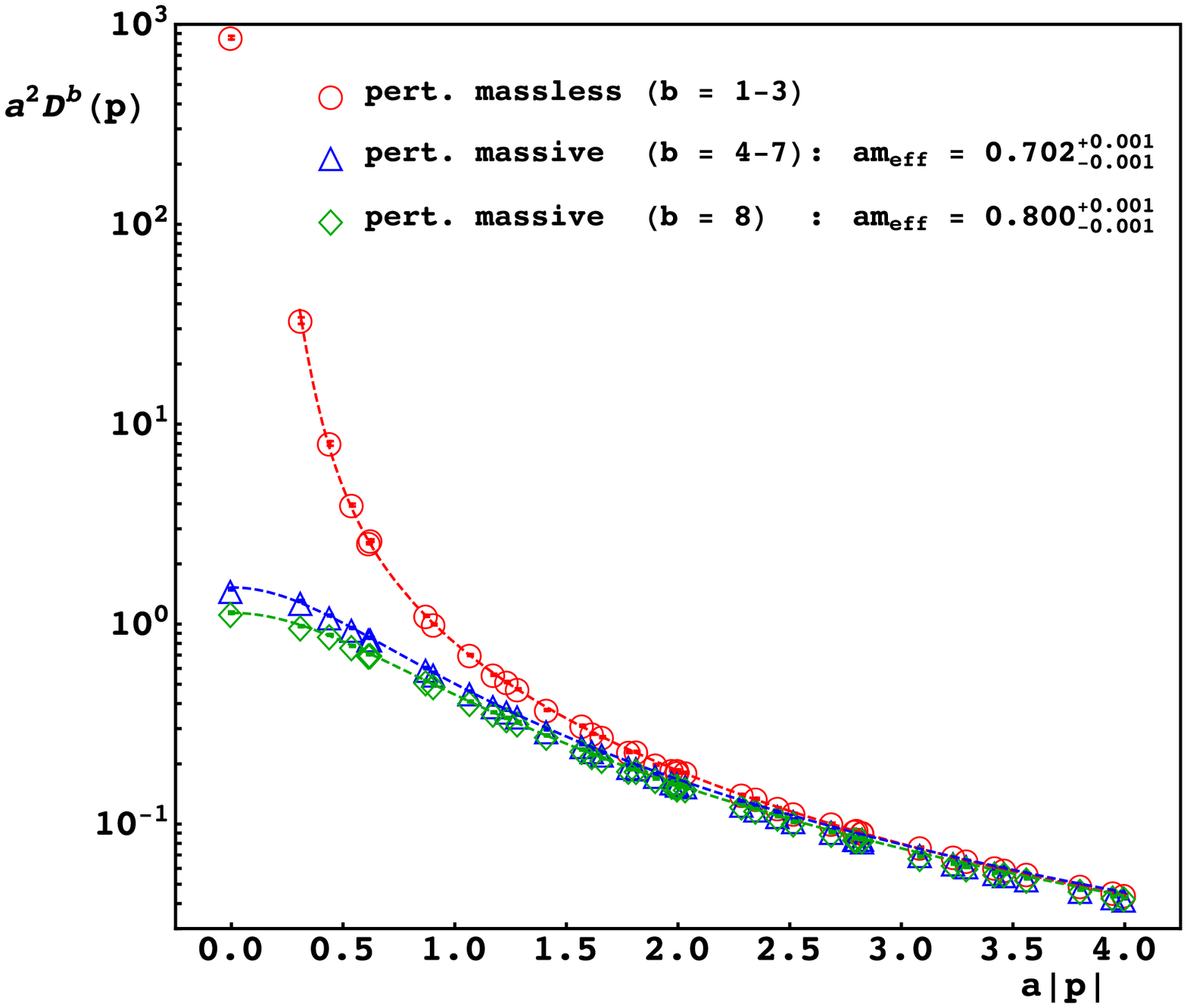}\\
\hspace{-0.25cm}
\includegraphics[width=0.455\textwidth]{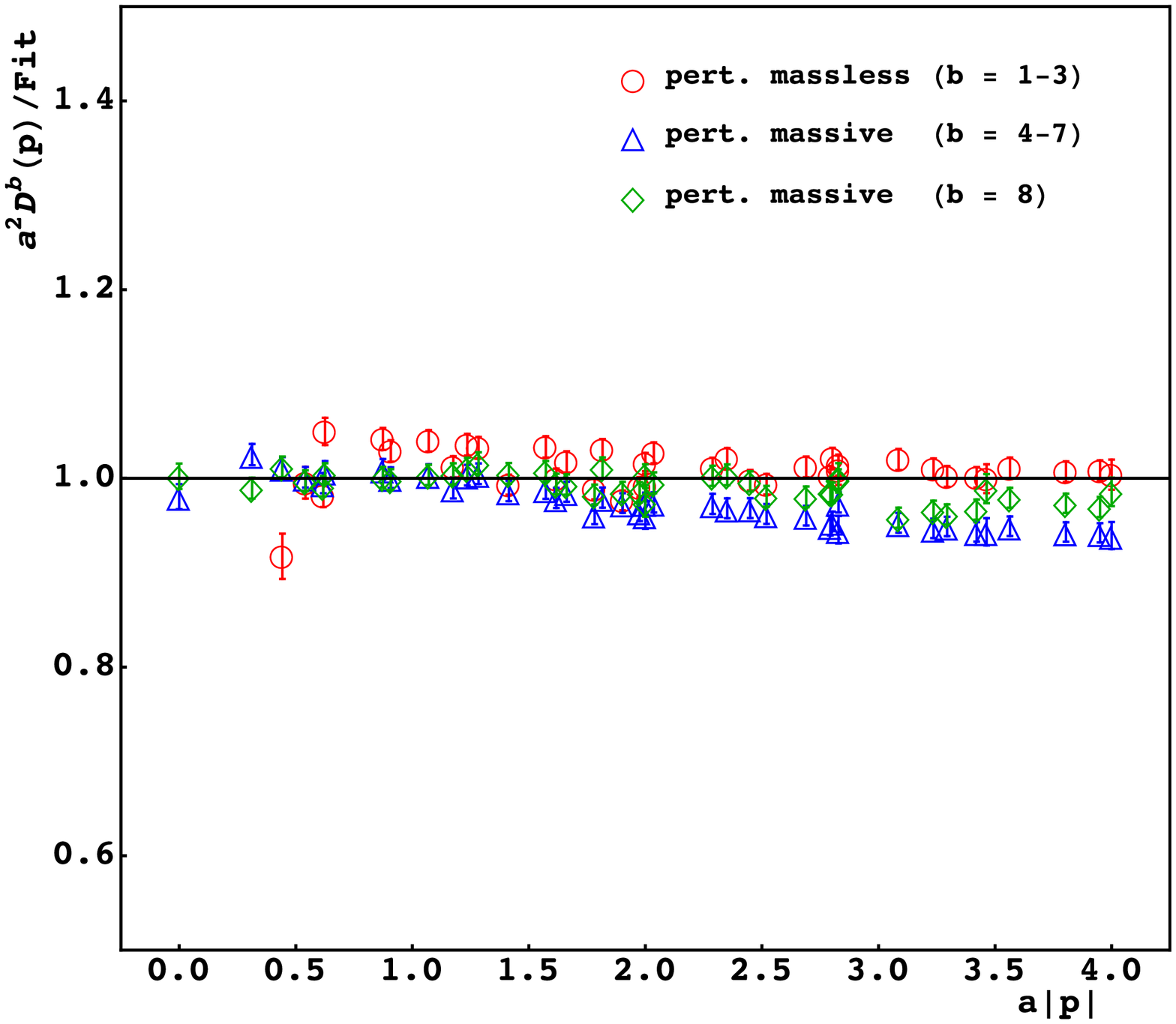}
\caption{Top: Propagators in the unbroken $SU(2)$ subgroup and the coset $SU(3)/SU(2)$ for a $V=20^4$ lattice. The momenta are along the edges, and along all possible diagonals of the lattice, i.e. $(p,0,0,0),\dots,(p,p,p,p)$. Results are for the parameters $\beta=9.59055$, $\kappa=0.444462$, $\lambda=0.4118$. The dashed lines denote the fits discussed in the text. Bottom: Differences of the lattice propagators and the fits.}
\label{fig:w}
\end{figure}
 
The results are shown in Figure \ref{fig:w} for a $V=20^4$ lattice. The propagators are very close to the expected perturbative behavior
\begin{eqnarray}
&&D_{\text{broken}}(p^2)= \frac{Z}{p^2+(am_{\text{eff}})^2}\;,\nn
\end{eqnarray}
\newpage
\begin{eqnarray}
&&D_{\text{coset}}(p^2)=\frac{Z}{p^2}\times\nn\\
&&\times\left(\frac{a}{Vp^4}+\frac{p^2}{(am_{\text{eff}})^2 + b^2 p^2 \left(1 + c^2 \ln\frac{\Lambda^2 + p^2}{\Lambda^2}\right)^\gamma}\right)\nn\;,
\label{eq:fit1}
\end{eqnarray}
\no where $Z$ are wave-function renormalization constants and $am_{\text{eff}}$ is the fitted mass in lattice units. Note that for the massive propagator the mass dominates sufficiently that the logarithmic corrections of leading loop corrections only play a minor role, as the small deviations at large momenta show, and are therefore ignored here. For the massless case, this effect is larger. Note that in this case the first term is a pure finite-volume effect, while a very small mass cannot be excluded yet, but is at least two orders of magnitudes smaller than the one for the massive propagators, and will therefore be ignored in the following. This also supports that the dynamics of the unbroken sector does not influence the broken sector, as the latter shows the expected behavior.

This close resemblance to the perturbative expectations shows that corrections to the propagators are perturbatively small, supporting the applicability of the FMS mechanism. This also implies that the gauge-dependent spectrum of the elementary particles coincides with the one expected in perturbation theory, especially of three massless and five massive states. The (almost) masslessness of the propagator in the unbroken sector suggests a Coulomb-like behavior, although corrections deep in the infrared  may still alter this behavior. The ratio between the masses of the $4+1$ gauge bosons in the coset is about $0.90(1)$ and the tree-level ratio is $\sqrt{3/4}\approx 0.87$ \cite{Torek:2015ssa,Maas:unpublishedtoerek}, again confirming the applicability of (tree-level) perturbation theory.

\begin{figure}[!t]
\centering
\hspace*{-1cm}
\includegraphics[width=0.54\textwidth]{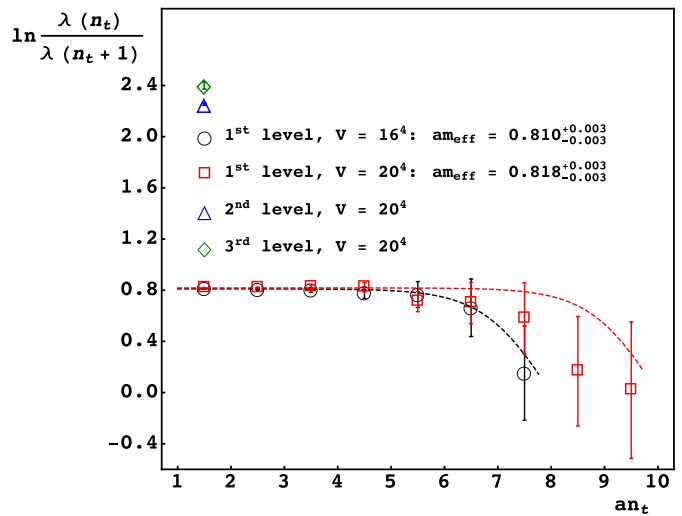}
\caption{Results from the variational analysis for the energy levels of the $J^{P}=1^{-}$ singlet channel are shown for the largest lattices ($V=16^4, 20^4$). The first three energy levels are shown and only the ground state gives a reliable result. Points with too large errors haven been suppressed.}
\label{fig:m}
\end{figure}

We also determined the effective mass using the Schwinger function as described in \cite{Maas:2013aia}. However, the statistics at this point is too low in order to make definite statements but we find within errors agreement with the masses extracted by the fits in momentum space.

To check the left-hand-side of Equation \pref{eq:fmsexp}, the next step is to perform the spectroscopy in the $1^-$ singlet channel. Since the number of degenerate states is an important prediction, it is not reasonable to just use a single operator. Rather a variational analysis \cite{Gattringer:2010zz,Maas:2014pba} using an operator basis is performed. This basis is constructed from the discretization of Equation \pref{eq:op},
\be
O_\mu=\text{Re}\left( i~\tr\left[\phi_i^\dagger U_\mu^{ij}\phi_j\nn\right]\right)\;.
\ee
\no In addition to this operator, a scattering state $O_\mu O_\nu O_\nu$ \cite{Maas:2014pba} is included in the basis. These operators are obtained from smeared gauge-field configurations to reduce noise \cite{Maas:2014pba}, where the gauge links were stout-smeared \cite{Morningstar:2003gk} and the Higgs fields APE-smeared \cite{Philipsen:1996af}. The final operator basis consisted twice of both operators, either 3-times or 4-times smeared. An extended basis will be discussed in \cite{Maas:unpublishedtoerek}.

The results for the level spectrum obtained from $100K$ configurations for the $V=16^4$ lattice and from $90K$ configurations for the $V=20^4$ lattice  are shown in Figure \ref{fig:m}. The effective mass of the ground-state was obtained from a single-cosh fit (dashed lines in Figure \ref{fig:m}). The mass of this state is about $am_{\text{eff}} \approx 0.82(1)$, in good agreement with the FMS prediction \pref{eq:fmsexp}. All other levels are too noisy for a final statement, but are very high up in the spectrum. In fact, the available points put them at $am_{\text{eff}} \gtrsim 2.4$, exactly where the first scattering state is expected to be if the scalar mass should be above the elastic threshold \cite{Maas:2014pba}. Thus, the degeneracy of the ground state is one, as predicted by the FMS mechanism.

\begin{figure}[!t]
\centering
\hspace*{-2cm}
\includegraphics[width=0.5\textwidth]{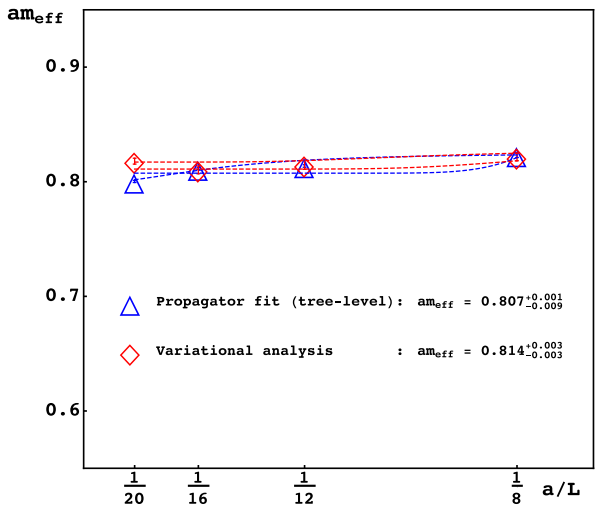}
\caption{Volume-dependence of the effective masses for the massive $W$-propagator and the $1^-$ channel ground-state. The dashed lines are fits 
to $am_{\text{eff}}=A+B e^{-cV}$.}
\label{fig:vol}
\end{figure}

Moreover, the volume dependence of the masses, shown in Figure \ref{fig:vol}, yield that both, the mass of the massive gauge boson propagator and the ground-state in the $1^-$ channel, are more or less constant, and in particular show no characteristic polynomial decay as expected for massless particles \cite{Gattringer:2010zz}.  Thus, in contradiction to the conventional perturbative expectation, there is no massless state in the singlet vector channel. Instead, the spectrum agrees with the results of gauge-invariant perturbation theory.

Of course, as suggestive as these results are, it is necessary to improve the systematic reliability by enlarging the operator basis, reducing statistical errors, and improving on systematic effects from discretization and volume. This is ongoing, and will be presented at a later time \cite{Maas:unpublishedtoerek}. However, the two predictions are so qualitatively different and the available results agree so rather well with one of them that it appears at the current time unlikely that any fundamental change of the results could be induced by systematic errors.

\section{Summary and conclusions}

Summarizing, we have shown that, in an $SU(3)$ gauge theory with one fundamental Higgs field the perturbative framework does not predict the correct spectrum. However, combining it with the FMS mechanism in gauge-invariant perturbation theory fixes this and predicts the correct spectrum, at least in the singlet vector channel discussed here. Still, the perturbative description works exceedingly well for the gauge boson propagators. Thus it is possible to predict, entirely on basis of gauge-invariant perturbation theory, the correct mass spectrum. This implies that while conventional perturbation theory may not be able to predict correctly the experimentally accessible spectrum, it is still an important component in the framework of gauge-invariant perturbation theory, which can.

Note that the question if a critical region in the parameter space of this particular theory exists, where the correlation length diverges, i.e. a region where the continuum limit can be taken, cannot yet be answered. Still this theory can be considered as a low energy effective theory\cite{Hasenfratz:1986za}, for which the observations made apply.

While at first sight very special, similar issues may or may not happen in other BSM scenarios \cite{Maas:2015gma,Maas:2016qpu}, and even small alterations in the theory, like adding a second Higgs field, can again change the picture completely \cite{Torek:2015ssa}. Also, dynamical effects can spoil even the possibility to combine perturbation theory and the FMS mechanism to predict the physical spectrum without full non-perturbative calculations \cite{Maas:2014pba}. These questions require further understanding to make a reliable and easy prediction of the physical spectrum possible for model building, and also to understand the limitations. \\

\section*{Acknowledgements}
We are grateful to L.\ Pedro for helpful discussions. P.\ T.\ has been supported by the FWF doctoral school W1203-N16. The computational results presented have been obtained using the Vienna Scientific Cluster (VSC) and the HPC center at the University of Graz.

\bibliographystyle{bibstyle}
\bibliography{bib}

\end{document}